\documentclass[a4paper,11pt]{article}
\usepackage{jheppub} 

\usepackage{epsfig}
\usepackage{graphicx}
\def\be{\begin{equation}}
\def\ee{\end{equation}}
\def\bea{\begin{array}}
\def\eea{\end{array}}
\def\beqa{\begin{eqnarray}}
\def\eeqa{\end{eqnarray}}
\def\beqas{\begin{eqnarray*}}
\def\eeqas{\end{eqnarray*}}

\def\bp{\begin{picture}}
\def\ep{\end{picture}}
\def\bc{\begin{center}}
\def\ec{\end{center}}
\def\bfig{\begin{figure}}
\def\efig{\end{figure}}

\def\bit{\begin{itemize}}
\def\eit{\end{itemize}}
\def\nn{\nonumber}
\def\f{\frac}

\def\[{\left[}
\def\]{\right]}
\def\({\left(}
\def\){\right)}

\def\..{\left.}
\def\.{\right.}
\def\tl{\tilde}
\def\ra{\rightarrow}
\def\la{\leftarrow}

\def\tm{\times}

\def\da{\dagger}

\def\la{\lambda}

\def\al{\alpha}

\def\ka{\kappa}

\def\ep{\epsilon}

\def\Ga{\Gamma}
\def\ga{\gamma}
\def\pa{\partial}
\def\pr{\prime}

\title{ Explain DAMPE Results by Dark Matter With Hierarchical Lepton-Specific Yukawa Interactions}

\author[a]{Guo-Li Liu ,}
\author[a,b]{Fei Wang,}
\author[c]{Wenyu Wang,}
\author[b,d,e]{Jin Min Yang}

\affiliation[a]{School of Physics, Zhengzhou University, Zhengzhou 450000, P. R. China}
\affiliation[b]{CAS Key Laboratory of Theoretical Physics, Institute of Theoretical Physics,
                Chinese Academy of Sciences, Beijing 100190, P. R. China}
\affiliation[c]{College of Applied Science, Beijing University of Technology, Beijing 100124, P. R. China}
\affiliation[d]{School of Physics, University of Chinese Academy of Sciences, Beijing 100049, P. R. China}
\affiliation[e]{Department of Physics, Tohoku University, Sendai 980-8578, Japan}

\emailAdd{guoliliu@zzu.edu.cn}
\emailAdd{feiwang@zzu.edu.cn}
\emailAdd{wywang@mail.itp.ac.cn}
\emailAdd{jmyang@itp.ac.cn}

\abstract{ We propose to interpret the DAMPE electron excess at 1.5 TeV through scalar
or Dirac fermion dark matter (DM) annihilation with doubly charged scalar mediators
that have lepton-specific Yukawa couplings.  Hierarchy of such lepton-specific Yukawa
couplings is generated through the Froggatt-Nielsen mechanism, so that the dark matter
annihilation products can be dominantly electrons. Stringent constraints from LEP2 on
intermediate vector boson production can be evaded in our scenarios.
In the case of scalar DM, we discuss one scenario with DM annihilating directly to leptons
and the other scenario with DM annihilating to scalar mediators followed by their decays.
We also discuss the Breit-Wigner resonant enhancement and the Sommerfeld enhancement in case
that the s-wave annihilation process is small or helicity suppressed. With both types of
enhancement, constraints on the parameters can be relaxed and new ways for model building
will be open in explaining the DAMPE results.
}

\begin{document}
\maketitle \indent
\newpage
\section{Introduction}
The nature of dark matter (DM) is one of the most important questions in particle physics and
cosmology. In fact, many new physics theories beyond the standard model (SM) can provide viable
DM candidates. Apart from its various gravitational influences, dark matter has so far eluded
all direct detection experiments through nucleon recoil, prompting ones to find alternative
ways to search for its existence. An important method of probing DM properties is via indirect
detections, whereby we look for the appearance of particles like high energy gamma-rays,
positrons or X-rays produced via annihilation or decay of DM concentration in galaxies
(satellites, dwarfs, or clusters).

In past few years, positron excess has been reported by various experiments, such as AMS02\cite{AMS1,AMS2},
PAMELA\cite{PAMELA1,PAMELA2} and Fermi\cite{FermiLAT1}.
Recently, the DArk Matter Particle Explorer (DAMPE) satellite\cite{Collaboration2017}, which is a new cosmic
ray detector with excellent energy resolution and hadron rejection power, published their measurements about
the cosmic $e^+ + e^-$ flux up to 5 TeV and announced to find a sharp peak at $\sim 1.5 \rm TeV$.  Although
both astrophysical (e.g., pulsars) and DM origins are possible to act as the required nearby mono-energetic
electrons sources, the DM explanation could potentially guide the search of DM particles in future direct
detection and collider experiments. Many DM models had been proposed to explain the DAMPE results
\cite{DAMPE-1,DAMPE-2,DAMPE-3,DAMPE-4,DAMPE-5,DAMPE-6,DAMPE-7,DAMPE-8,DAMPE-9,DAMPE-10,DAMPE-11,DAMPE-12,DAMPE-13,DAMPE-14,DAMPE-15,DAMPE-16, DAMPE-17,DAMPE-18,DAMPE-19,DAMPE-20,DAMPE-21,DAMPE-22,DAMPE-23,DAMPE-24,DAMPE-25,DAMPE-26,DAMPE-27,DAMPE-28,DAMPE-29}. In the DM explanation (DM annihilation into electrons
or equal amounts of lepton flavors), the best fit value of the DM mass should be 1.5 TeV if
the nearby DM sub-halo locates at $\rm 0.1 \sim 0.3 kpc$  away form the solar system and the DM annihilation
cross section should be $\langle\sigma v\rangle\approx3\tm 10^{-26}{\rm cm}^3/s$. Besides, in order to guarantee
the resulting electron/positron spectrum to be a narrow peak instead of a box shape spectrum, the required mass
ratio between the mediator (to lepton pairs) and the DM mass is stringently constrained to be higher than
0.995 by numerical fitting to the DAMPE data \cite{DAMPE-7}.

In order to have a large DM annihilation cross section in the DM sub-halo and at the same time give the
correct DM relic density, it is preferable to adopt Dirac DM scenarios because the Dirac DM annihilation
into lepton pair final states via vector leoptophilic mediator will not be s-wave suppressed\cite{luty}.
Scalar DM scenarios, which can annihilate into vector mediator pairs followed by their late-time decays,
can also explain the result. Both scenarios tend to adopt a leptophilic gauge boson as the mediator which
can couple universally to the lepton flavors.
On the other hand, the DM direct detection bounds as well as the electron/positron collider constraints on
the vector mediator production will impose rather stringent constraints on these scenarios.

An alternative possibility is the scalar-portal DM scenario, in which the mediator will mainly couple
to leptons. We propose to explain the DAMPE excess by a scalar or Dirac fermion DM candidate with
lepton-specific Yukawa couplings. A horizontal family symmetry $U(1)_H$, which can be either global
(with small explicit symmetry breaking terms) or local, is introduced for the lepton sector with
Froggatt-Nielsen mechanism\cite{FN-1,FN-2}. By properly choosing the $U(1)_H$ quantum numbers, the Yukawa
couplings between the scalar mediators and the first family leptons can be unsuppressed while other types
may be suppressed, which is just needed for explaining the DAMPE result.

In general, the models with s-wave suppressed DM annihilation cross section are not favored
to interpret the DAMPE results.
We note that, by adopting Breit-Wigner enhancement\cite{BW-1,BW-2,BW-3} or Sommerfeld enhancement\cite{SO-1,SO-2,SO-3}, the s-wave (helicity or propagator) suppressed process can also be used to explain the DAMPE results
because of a large enhancement factor. Therefore, new ways for model building in explaining the DAMPE results
will be open with such an enhancement factor.

This paper is organized as follows. In Section~\ref{section2}, we introduce a scalar DM model to explain the
DAMPE electron/positron excess. Effects of Breit-Wigner enhancement and Sommerfeld enhancement are discussed.
In Section~\ref{section3}, a model with Dirac fermion DM is proposed.  Finally, we draw our conclusions in
Section \ref{section4}.

\section{ Scalar dark matter with scalar mediator involving lepton-specific interactions
} \label{section2}

We propose to explain the DAMPE result with scalar DM and scalar mediators. The complex scalar $S$, which is
odd under a discrete $Z_2$ symmetry, will act as the DM candidate while other fields are even under $Z_2$.
A Higgs-portal type interactions between $S$ and scalar mediator $T$, which will couple only to leptons,
will be introduced. To generate the Yukawa hierarchy so that the mediator decays dominantly to electrons,
a global $U(1)_H$ with small explicit symmetry breaking terms or an anomaly free local $U(1)_H$ horizontal
symmetry will be introduced to generate the required suppression factors via Froggatt-Nielsen mechanism.
The Lagrangian has the following form
\beqa
{\cal L}\supseteq&& |\pa_\mu S|^2+|D_\mu T|^2-m^2_S |S|^2-m^2_T |T|^2-\f{\la_1}{4}|S|^4-\la_2|S|^2|T|^2-\f{\la_3}{4}|T|^4~ \nn\\
 && -y_i\sum\limits_{i}\(\f{U}{\Lambda}\)^{Q^E_{ij}} \(\bar{E}_{R,i}^c E_{R,j} T\)+\cdots~,
\eeqa
where
\beqa
D_\mu T=(\pa_\mu-i Q_Y^T g_Y  B_\mu) T~, ~~Q_{ij}^E\equiv -\f{Q_H(T)+Q_H(E_R^i)+Q_H(E_R^j)}{Q_H(U)}~,
\eeqa
with $Q_Y^T=2$.

The non-renormalizable interactions involving the flavon field $U$, which transform non-trivially under
$U(1)_H$, are generated after integrating out the heavy modes at the scale $\Lambda$. Due to the charge
assignments of $U(1)_H$ horizontal symmetry for the SM fermions and $T$, the Yukawa coupling of the form
$\bar{E}_{R,i}^c E_{R,i} T$ will appear only after the flavon acquires a VEV $\langle U\rangle$.

 \begin{table}[h]
\caption{The local horizontal $U(1)_H$ quantum numbers for the SM matter contents with the generation index
$a=1,2,3$. }
\centering
\begin{tabular}{|c|c|c|c|c|c|c|c|c|c|}
\hline
&$Q_L^a$ &$U_R^a$&$D_R^a$& $L_{L,e},e_R$&~$L_{L,\mu},\mu_R$&$L_{L,\tau},\tau_R$&S&T&U\\
\hline
$U(1)^\pr$&0&0&0&-2&~1&~1&~0&4&-1\\
\hline
\end{tabular}
\end{table}

From the $U(1)_H$ quantum numbers given in Table I, the Yukawa couplings between the scalar mediator and
leptons have the following hierarchy
\beqa
{\cal L}\supseteq - y_1\bar{E}_{R,i}^c E_{R,i} T-y_2\(\f{U}{\Lambda}\)^6 \(\bar{\mu}_{R}^c \mu_{R,i} T\)-y_3\(\f{U}{\Lambda}\)^6 \(\bar{\tau}_{R}^c \tau_{R,i} T\)~.
\eeqa
After $U$ acquires a VEV so that $\langle U\rangle/\Lambda \sim {\cal O}(0.1)$,
for $y_1\simeq y_2\simeq y_3\sim {\cal O}(1)$, the resulting Yukawa couplings have the following hierarchy
\beqa
y_e\gg y_\mu\approx y_\tau~.
\eeqa
An alternative possibility is that we choose the following $U(1)_H$ quantum numbers
\beqa
Q^\pr(e,\mu)=1,~Q^\pr(\tau)=-2~,Q(T)=-2~,Q(U)=1.
\eeqa
The Yukawa couplings for $e,\mu$ are unsuppressed while for $\tau$ is suppressed,
so $y_e\simeq y_\mu\gg y_\tau$. Similarly, we could have $y_e\simeq y_\tau\gg y_\mu$.

For the DM annihilation, we have two possibilities
\bit
\item {\bf Scenario I}:

The DM particles annihilate into the scalar mediators $S S^*\ra T^{++} T^{--}$, followed by their
decays into lepton pairs $T^{--}\ra l_R l_R$.
So the dark matter mass should satisfy $m_\chi\equiv m_S> m_{T}$ and the range of
$m_T\in [0.995, ~1]\times m_S$ to fit the peak shape eletron/positron spectrum required to fit the DAMPE data.

\begin{figure}[htb]
\begin{center}
\includegraphics[width=4.0in]{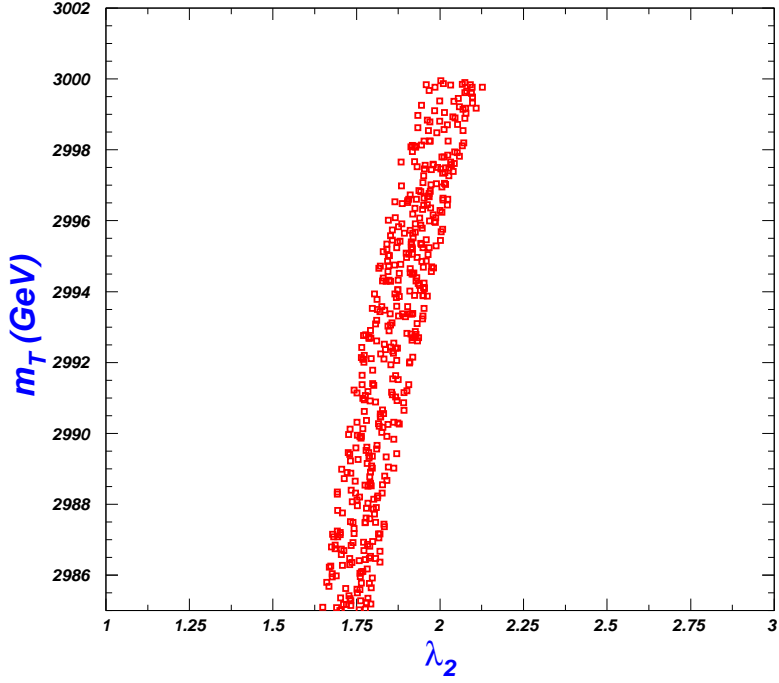}
\end{center}
\vspace{-.5cm}
\caption{ The range of the parameter $\la_2$ vs $m_T$ to explain the DAMPE result without enhancements
in the case of scalar DM.}
\label{fig1}
\end{figure}

Our numerical results are shown in Fig.\ref{fig1}.
We can see that the Higgs portal coupling $\la_2$ is constrained
to lie near 1.7, which is quite large but within the perturbative regime.

In previous papers with typical leptophilic $U(1)$ gauge bosons as DM mediator\cite{DAMPE-12,DAMPE-21},
the LEP2 bounds on vector boson production will rule out many simplest scalar DM models. In our setting,
the LEP2 constraints will be relaxed because the $T$ scalar with double electric charge will not be produced
directly on the electron/positron collier LEP2. So it is fairly advantageous to use such doubly charged scalar
as the DM mediator.

Naively, the DM can scattering off the light quarks at one-loop order through $T$ scalar loops that connect
to the quark lines via photons. However, the one-loop diagram can be proved to vanish. So the leading order
contributions arise at two loop order.  The DM direct detection cross section are thus suppressed at two-loop
order. By matching to the effective operator after integrating out the scalar $T$
  \beqa
  {\cal L}_{eff}\approx \f{2 }{3}\al_{em}^2 \la \f{\chi^\da\chi}{m_T^2}\f{F^{\mu\nu}F_{\mu\nu}}{e^2}~,
  \eeqa
we can estimate the scalar DM-nucleon scattering amplitudes to be
\beqa
{\cal M}\approx \f{\ka}{12}\al_{em}^2 \f{\la}{m_T^2}Z^2\tl{F}(q^2)\(\bar{u}_N^\pr\f{1}{2}(1+\ga^0)u_N)\).
\eeqa
Following the methods in \cite{0907.3159}, we could estimate the DM direct detection cross section
\beqa
\f{d \sigma}{d E_d}\approx \(\f{\al_{em}Z}{\pi}\)^2\f{m_N}{2\pi v^2}\(\f{\al_{em} Z}{\pi} \lambda \)^2\(\f{\pi^2}{12}\)^2\f{m_N^2}{m_T^2}v_d^2\tl{F}(q^2)~,
\eeqa
with the DM velocity $v\sim 10^{-3}c$ and  the velocity of the recoiled nucleus $v_d=\sqrt{2 E_d/m_N}$.
With $m_T\sim 3 TeV$, the DM-nucleus scattering cross section, which can be estimated to be of order
$3\times10^{-16} pb$, can easily survive the DM direct detection bounds given by LUX\cite{LUX2016} and
PandaX\cite{PANDAX}.

\item {\bf  Scenario II}:

The DM particles annihilate directly into four leptons $S^* S\ra  (T^{++})^* (T^{--})^*\ra l^-l^- l^+l^+$
if the scalar mediator is heavier than 3 TeV. Such  annihilation cross section will in general be suppressed
by the mediator propagator.

\eit
     In order to enhance the DM annihilation cross section to explain the DAMPE data, we propose two ways
in this scenario.
     \bit
     \item {\bf Resonant annihilation with Breit-Wigner enhancement}:

   If the scalar mediator mass lies slightly higher than the DM mass $m_S$, the direct 4-lepton final states
annihilation will be enhanced by Breit-Wigner resonant effect\cite{BW-1}.
   The resonant annihilation cross section are given as
   \beqa
   \sigma v_{\rm rel}&\approx& \f{\la_{2}^2}{4m_\chi^2 }\[\f{2m_T \Ga_T}{(p^2-m_T^2)^2+m_T^2\Gamma_T^2}\]^2~,\nn\\
   &\approx& \f{\la_{2}^2}{ 4m_\chi^2 } \[\f{8 \gamma }{(v^2_{rel}-2\epsilon)^2+4\ga^2}\]^2~,
   \eeqa
       with
       \beqa
       && p^2\approx m_\chi^2+\f{1}{2}m_\chi v_{rel}^2~,
~~ \epsilon=\f{m_T^2-m_\chi^2}{m_T^2}~,~~~~~~\gamma=\f{m_T\Ga_T}{m_T^2}.
       \eeqa
     It is known that for $\ga\ll 1$ and $\ga^2\ll(v^2_{rel}-2\epsilon)^2$, we have the approximation
     \beqa
    \lim\limits_{\ga \ra 0} \f{8 \gamma }{\[(v^2_{rel}-2\epsilon)^2+4\ga^2\]} = 4\pi \delta(v^2_{rel}-2\epsilon)~.
     \eeqa
   We can see that we could have {\bf double} enhancement near the threshold due to the apparent $\delta(0)$.
We need to cut the apparent infinity with the decay width of $T$ which is estimated to be given by
   \beqa
   \Ga_T\approx \f{ y_E^2 m_T}{32\pi}~.
   \eeqa
    So we have the cut for the delta function
    \beqa
    \delta(v^2_{rel}-2\epsilon)\ra  \f{m_T}{\pi \Ga_T}\approx \f{32}{y_E^2}.
    \eeqa
  Using the standard formula for DM relic density\cite{BW-3},  $\Omega h^2$ is given by
   \beqa
   \Omega h^2\approx 2.755 \tm 10^8 \sqrt{\f{45}{\pi g_*}}\f{1}{M_{pl} J_f} {\rm GeV}^{-1}~,
   \eeqa
    with
    \beqa
    J_f&=&\int\limits_{0}^\infty d v_{rel}  \f{v^2_{rel}(\sigma v_{rel})}{2\pi^{1/2}} \int\limits_{x_f}^{\infty} dx x^{-1/2} e^{-x^2 v_{rel}^2/4}~,\nn\\
    &\approx & \f{4\pi^2\la_2^2}{m_\chi^2} {\rm erfc}(\sqrt{\f{x_f \epsilon }{2}}) \delta(0)~,\nn\\
    &\approx& \f{4\pi^2\la_2^2}{m_\chi^2}{\rm  erfc}(\sqrt{\f{x_f \epsilon }{2}}) \f{32}{y_E^2}.
    \eeqa
So we can obtain a universal $\langle\sigma v\rangle$ at the freeze-out time. The thermal averaged cross
section today can also be readily obtained from $J_f$ with $\epsilon\ra 0$. So, if this scalar DM scenario
is to explain the DAMPE data, the range of $\la_2$ can be much smaller if $y_E$ is small because of the
double enhancement effects near resonance. We should note that the resonant enhancement will affect the
annihilation cross section at both the freeze-out temperature and today. The DM-nucleon scattering cross
section will be suppressed by a tiny coupling $\lambda_2$, so the DM direct detection experiments may not
see any signal in the near future.

     \item {\bf DM annihilation with Sommerfeld enhancement}:
\begin{figure}[htb]
\begin{center}
\includegraphics[width=4.0in]{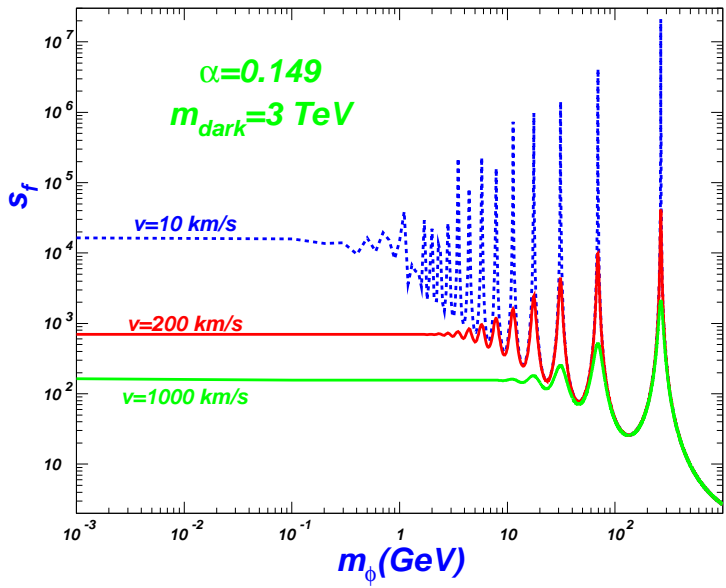}
\end{center}
\vspace{-1.0cm}
\caption{The Sommerfeld enhancement factor $S_f$ vs the light gauge boson mass $m_\phi$
for $\al_X \approx 0.149$ and DM mass $m_{S}=3 {\rm TeV}$. Here $v=10 {\rm km/s},200 {\rm km/s},1000 {\rm km/s}$ correspond to the characteristic speeds of dwarf halos, the Milky Way and clusters, respectively.
}
\label{fig2}
\end{figure}
There is an alternative possibility in which the DM annihilation in the early universe may be helicity
suppressed (by light fermion mass) at the s-wave and the DM relic density is determined mainly by p-wave
processes.  Suppressed by the low relative velocity $v\sim 10^{-3} c $, the p-wave process will not give
dominant contributions to the DM annihilation in the sub-halo. So, in order to explain the observed
electron/positron excess reported by DAMPE, we need to enhance the suppressed s-wave contributions
to the DM annihilation today.

Fortunately, the Sommerfeld enhancement has the desired behavior: it is enhanced at low velocities and
therefore can boost the present day annihilation cross section. The DM annihilation in the early universe
will be affected only slightly. We should note although chemical recoupling\cite{SO-3} implies suppressed
indirect detection signals for near resonance regions, the Sommerfeld enhancement can still enhance the
suppressed annihilation up to a factor $10^2\sim 10^3$ to account for the DAMPE signal.

In this scenario, although the s-wave of DM annihilation is not suppressed, the requirement to give
a correct DM relic density will stringently constrain the couplings involved. Considering the Sommerfeld
enhancement can explain the DAMPE results with much smaller coupling strengths. Therefore, the DM direct
detection bounds can be evaded very easily.

We can introduce an additional $U(1)_X$ gauge group with the gauge coupling strength $g_X$ and the gauge
boson mass $m_X$. The complex DM particle $\chi\equiv S$ will transform non-trivially under $U(1)_X$ with
the corresponding charge $Q_S$.
The Sommerfeld enhancement factor can be approximated by\cite{SO-2}
      \beqa
     \tl{S} \approx \f{g_X^2 m_S}{4\pi m_X}~.
      \eeqa
for a massive gauge boson mediator $A_\mu^X$. While the DM annihilation is dominantly
by $S^* S\ra XX$ via $t,u$-channel $S$,  the resulting cross section
      \beqa
      \langle \sigma v_{rel} \rangle\approx \f{ g_X^4 Y_S^4}{48\pi m_S^2}~
      \eeqa
must give the correct DM relic density,  which is given by the Planck satellite
data $\Omega_{DM} = 0.1199\pm 0.0027$ \cite{Planck}
in combination with the WMAP data \cite{WMAP}(with a $10\%$ theoretical uncertainty).
So we must have $ \al_X Y_S^2\approx 0.149$ for the DM mass $m_{S}=3 {\rm  TeV}$.
For simplicity, we choose the DM $U(1)_X$ charge $Y_S=1$. Numerical results on the value of
Sommerfeld enhancement factor are given in Fig.\ref{fig2}. We can see that, given an appropriate mass of the mediator $\phi$,
sufficient enhancement factor can be achieved at different cosmic objects, such as the dwarf halos, Milky Way and the clusters.

It is addressed in \cite{SO-3} that the DM can annihilate into  $A_\mu^X \phi$ with $\phi$ being the hidden
Higgs that breaks the $U(1)_X$ gauge symmetry. The mass of $\phi$ is argued to be $m_\phi\lesssim 10 m_X$ by
perturbative requirement. However, we can introduce additional Higgs portal terms, which can contribute to
the mass of $\phi$, so that the simple relation between $m_\phi$ and $m_X$ is released. So we will neglect
the effect of hidden Higgs $\phi$ in our estimation of  $\langle \sigma v_{rel} \rangle$  by tuning $\phi$ to
be very heavy. Openning both annihilation channels will introduce uncertainty for the parameter $g_X$
because of the free parameter $m_\phi$.

\begin{figure}[htb]
\begin{center}
\includegraphics[width=4.0in]{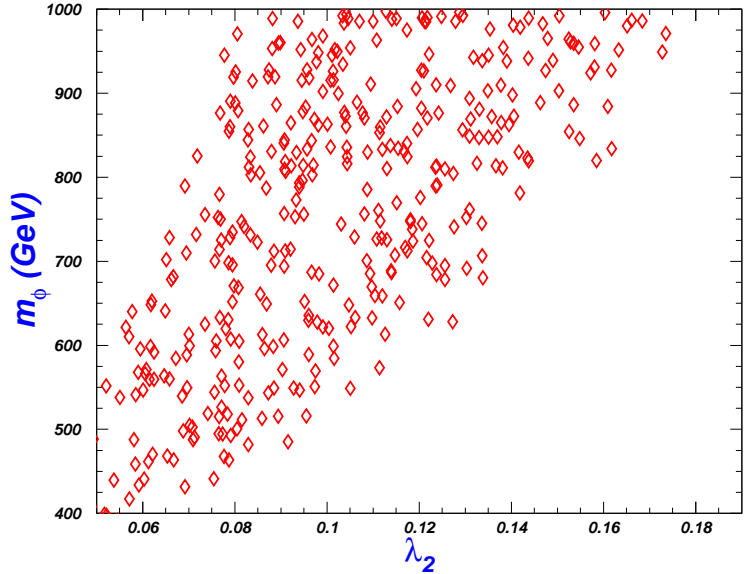}
\end{center}
\vspace{-.5cm}
\caption{ The allowed parameter $\la_2$ versus the light gauge boson mass $m_\phi$ when the Sommerfeld
enhancement is considered in the case of scalar DM.
}
\label{fig3}
\end{figure}
The Sommerfeld enhancement factor is determined by the DM wave-function obtained from non-relativistic
Schrodinger equation:
     \beqa
     \tl{S}=\left|\f{\chi(\infty)}{\chi(0)}\right|^2~,
     \eeqa
with which the approximate analytic results can be obtained after approximating the Yukawa potential by Hulthen potential\cite{SO-4}. The enhancement factor is thus approximated by
     \beqa
     \tl{S}\approx \f{\pi^2 \al_X m_X}{6 m_\chi v^2}
     \eeqa
at the resonance  and is proportional to $v^{-1}$ away from resonance. In order to avoid the chemical
recoupling, we will adopt mainly the $v^{-1}$ enhancement behavior in our numerical study.
In Fig.\ref{fig3} we can see that the coupling strength $\la_2$ can be reduced to a much smaller value with
such an enhancement.
As the DM annihilation cross sections into leptons are enhanced by the Sommerfeld factor, the reduced
coupling strengths of $\lambda$ will easily pass the DM direct detection constraints.
     \eit

\section{ Dirac fermion dark matter with scalar mediator involving lepton-specific interactions}\label{section3}
We also propose to explain the DAMPE results by introducing a Dirac DM scenario which can annihilate
into leptons via a scalar portal. It is also possible to adopt the s-wave suppressed Majorana DM scenario
with Sommerfeld enhancement and we will discuss that possibility in our subsequent studies.

The Lagrangian with Yukawa hierarchy from Froggatt-Nielsen mechanism is written as
\beqa
{\cal L}&\supseteq& \f{1}{2}(\pa_\mu S)^2+|D_\mu T|^2-\f{1}{2}m^2_S S^2-m^2_T |T|^2-\la S^2|T|^2-{\la}_2S^4-\la_3|T|^4-a_1 S |T|^2~ \nn\\
 &+&  i\bar{\chi} {\gamma}^\mu D_\mu \chi-m_\chi \bar{\chi}\chi- y_\phi\bar{\chi} \chi S
-\sum\limits_{a} y_a\(\f{U}{\Lambda}\)^Q \(\bar{E}_{R,a}^c E_{R,a} T\)+\cdots~,
\eeqa
with Dirac DM $\chi$ and an additional real scalar $S$ as a portal to doubly charged complex scalar $T$
that couple to the lepton species. The discrete $Z_2$  parity is imposed only for the DM particle $\chi$.

Similar to the preceding section, we can have different annihilation modes, depending on the mass of $S$
and $T$:
\begin{figure}[htb]
\begin{center}
\includegraphics[width=4.0in]{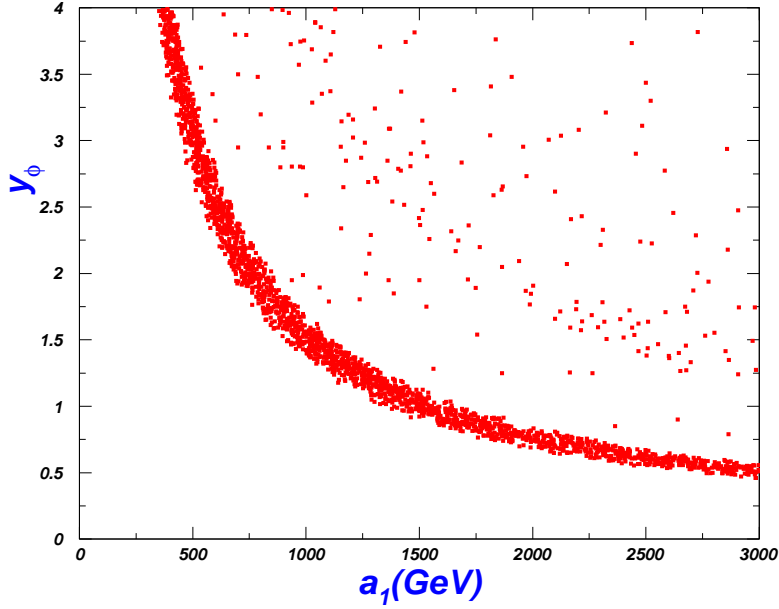}
\end{center}
\vspace{-.8cm}
\caption{ The allowed parameter space for $a_1$ vs $y_\phi$ to explain both the DAMPE results and
DM relic density in the case of Dirac DM without Breit-Wigner or Sommerfeld enhancement.
}
\label{fig4}
\end{figure}

\begin{figure}[htb]
\begin{center}
\includegraphics[width=4.0 in]{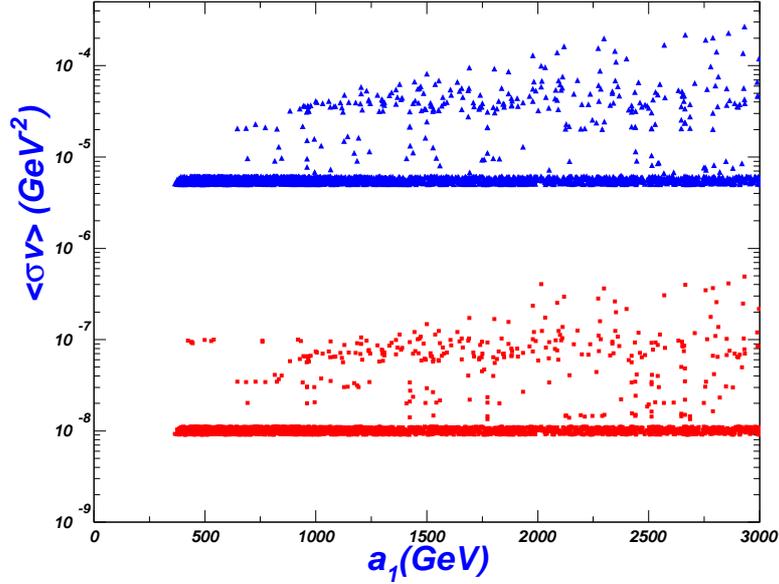}
\end{center}
\vspace{-.9cm}
\caption{ The scattering plot of the allowed parameter space for $a_1$ vs $\langle\sigma v\rangle$ to explain
both the DAMPE results and DM relic density in the case of Dirac DM with (blue) and without (red) Sommerfeld
enhancement. The $U(1)_X$ gauge boson is fixed to be $m_\phi=10 {\rm GeV}$.
}
\label{fig5}
\end{figure}

\bit
\item  For $m_S\simeq 6$ TeV and $m_T\in[0.995,~1]\tm m_\chi$, the DM can annihilate directly into $T T^*$ pairs.
This case is similar to the scenario I of scalar DM scenario. A new feature here is that resonant
enhancement annihilation is possible when $m_S$ is very near $2 m_\chi$. The discussions are similar to
that of scenario II in preceding section except that the $m_S$ can be slightly lighter than $2m_\chi$,
resulting in a negative $\epsilon$ parameter. In this case, the reduced coupling strength $\la$ can
still be able to give the correct DM relic density and at the same time explain the DAMPE results.
Numerical results on the parameters $a_1, y_\phi$, which can account for both the DAMPE data and
DM relic density, are given in Fig.\ref{fig4}.

\begin{figure}[htb]
\begin{center}
\includegraphics[width=4.0 in]{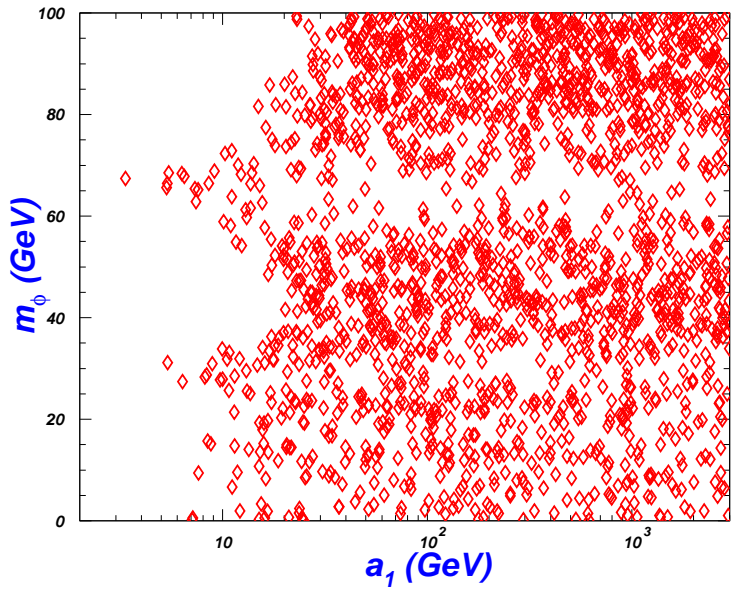}
\end{center}
\vspace{-.99cm}
\caption{ The allowed parameters $a_1$ vs the light gauge boson mass $m_\phi$ when
Sommerfeld enhancement factor is introduced in the case of Dirac DM.
}
\label{fig6}
\end{figure}

\item For $m_S\simeq 6$ TeV and  $m_T\gtrsim m_\chi$, the DM particles will annihilate directly into four
leptons $\bar{\chi}\chi\ra S\ra (T^{++})^* (T^{--})^*\ra l^-l^- l^+l^+$.
This case is similar to the correponding discussion in the preceding section. Both Sommerfeld enhancement
and resonance can be used to avoid the propagator suppression. Due to the fact that three internal propagators
exist, we could have triple enhancement due to the resonance. Sommerfeld enhancement are also possible by
introducing an additional light scalar or gauge boson.  So it is possible that the s-wave suppressed DM
annihilation cross section could be enhanced enough to explain the DAMPE results. Besides, the reduced
coupling strength $\la$ can be also possible with a large ( mass dimension) trilinear coupling $a_1$.
Fig.\ref{fig5} shows the the scattering plot of $\langle\sigma v\rangle$ with/without Sommerfeld enhancement
in the case of Dirac fermion DM. The allowed range of $a_1$ versus the light gauge boson mass $m_\phi$
with Sommerfeld enhancement is shown in Fig.\ref{fig6}.

\eit

\section{Conclusions} \label{section4}
In this work we proposed to interpret the DAMPE electron excess at 1.5 TeV through scalar (or Dirac fermion)
dark matter (DM) annihilation with doubly charged scalar mediators that have lepton-specific Yukawa couplings.
Hierarchical lepton-specific Yukawa couplings are generated through Froggatt-Nielsen mechanism, so that
the dark matter annihilation products can be dominantly electrons. Stringent constraints from LEP2 on
intermediate vector boson production can be evaded in our scenarios.
 In the case of scalar DM, we discussed two scenarios: one scenario with DM annihilating directly to
leptons and the other one with DM annihilating to scalar mediators followed by their decays.
We also discussed the Breit-Wigner resonant enhancement and  Sommerfeld enhancement in case that
the s-wave annihilation process is small or helicity suppressed. With both types of enhancement,
the constraints on the parameters can be relaxed and new ways for model building will be open
in explaining the DAMPE results.

\begin{acknowledgments}
This work was supported by the
Natural Science Foundation of China under grant numbers 11105124, 11105125, 11375001, 11675147, 11675242,
by the Open Project Program of State Key Laboratory of
Theoretical Physics, Institute of Theoretical Physics, Chinese Academy of Sciences
(No.Y5KF121CJ1), by the Innovation Talent project of Henan Province under grant
number 15HASTIT017 and the Young-Talent Foundation of Zhengzhou University,
by the CAS Center for Excellence in Particle Physics (CCEPP),  by the CAS Key Research
Program of Frontier Sciences and by a Key R\&D Program
of Ministry of Science and Technology of China under number 2017YFA0402200-04.

\end{acknowledgments}

\end{document}